\documentclass[prx,twocolumn,groupedaddress]{revtex4-2}

\usepackage{amsmath,amsfonts,amssymb,graphicx,xcolor,ulem}

\begin{document}


\title{ Multi-point distribution for Gaussian non-equilibrium non-Markovian observables }


\author{Roland R. Netz}
\affiliation{Fachbereich Physik, Freie Universit\"at Berlin, 144195 Berlin, Germany}
\affiliation{Centre for Condensed Matter Theory, Department of Physics, Indian Institute of Science, Bangalore 560012, India}


\date{\today}

\begin{abstract}
When analyzing  experimental or simulation time-series data, the question arises whether it is possible to tell  from  a one-dimensional time-dependent  trajectory whether the system is in equilibrium or not. We here consider the non-equilibrium version of the generalized Langevin equation for a Gaussian observable  and show that  i) the multi-point joint distribution  solely depends on the two-point correlation function and that ii) the two-point correlation function for a non-equilibrium process is identical to an equilibrium process with uniquely determined parameters. Since the multi-point joint distribution  completely characterizes the dynamics of an observable, this means that the non-equilibrium character of a system, in contrast to its non-Markovianity,  cannot be read off from a one-dimensional trajectory.

  \end{abstract}


\maketitle



The field of non-equilibrium  (NEQ) statistical mechanics
has steadily developed  for  decades
\cite{Prigogine1947,Mazur1953,Lebowitz1959,zwanzig_ensemble_1960,deGroot,grabert_microdynamics_1980,
Risken,zwanzig_nonequilibrium_2001,Jarzynski2000,Hatano2001,Seifert2005,Prost2009,Wynants2009,BroederszReview}.
In this development, projection techniques have played a key role,
since they allow to derive exact equations of motion, 
  so-called generalized Langevin equations (GLEs),
for arbitrary observables of interest
\cite{nakajima_quantum_1958,zwanzig_memory_1961,mori_transport_1965,Ciccotti1981,Straub_1987,lange_collective_2006,
kinjo_equation_2007,darve_computing_2009,hijon_morizwanzig_2010,izvekov_microscopic_2013,lee_multi_2019,
Ayaz2022,vroylandt_likelihood_2022}. It turns out that GLEs  contain, 
besides a deterministic force term that ensures approach to the steady-state  distribution in the long-time limit,
memory-friction  and complementary-force terms which account for dissipation and 
fluctuation effects, respectively. 
 The extraction of  all GLE parameters from simulation or experimental time-series data is
 well-established 
 \cite{Straub_1987,carof_two_2014,jung_iterative_2017,daldrop_external_2017,daldrop_butane_2018,
 klippenstein_cross-correlation_2021,vroylandt_likelihood_2022,Ayaz2022}. 
GLEs thus furnish a practical   
 connection between  fundamental concepts of  statistical mechanics and  experimental time-series  data, 
 the latter becoming progressively  available through 
 novel experimental approaches such as single-molecule and high-resolution optical techniques
\cite{Schmidt07,Gawedzki2009,Bechinger2010,Bocquet2012,Dinis2012,Bohec2013,Weitz2014}. 
In the past,   GLEs were amply used to study  protein folding dynamics 
\cite{ plotkin_non-markovian_1998,satija_generalized_2019,ayaz_non-markovian_2021,Dalton_2023},
barrier crossing dynamics
\cite{Bagchi_1983,Straub_1986,Pollak_1989,Carlon2018,Brunig_2022d},
the motion of living cells \cite{Mitterwallner_2020},
spectroscopy \cite{Tuckerman1993,Gottwald2015,Brunig_2022a},
dynamical networks \cite{Sollich2020} 
and  data prediction \cite{chorin_optimal_2000}.

The standard GLEs describe the time evolution  of an observable for an equilibrium (EQ) system 
and in particular obey the fluctuation-dissipation theorem (FDT)
\cite{nakajima_quantum_1958,zwanzig_memory_1961,mori_transport_1965},
hence   they do not apply
to driven i.e. stationary NEQ  systems, as described by a time-dependent Hamiltonian. 
A number of   works
dealt with generalizations of the  projection framework to time-dependent and
transient scenarios \cite{Robertson1966,Zwanzig1975,Picard1977,Uchiyama1999,Koide2002,Latz2002,
meyer_non-stationary_2017,Cui2018,Vrugt2019,meyer_non-markovian_2020},
but the resulting GLEs are typically too complex to be dealt with in practical applications. 
This is why in applications to driven NEQ systems, it has become standard practice to use
GLEs that break the FDT  more or less by hand \cite{Carlon2018,Lavacchi2022}.
The same starting point is taken in this paper. We consider a general GLE as  derived by
Mori projection and choose a random force distribution that is not related to the memory-friction  kernel
according to the FDT.  We ask whether from the time-dependent  trajectory of a Gaussian observable
we can tell whether the data come from a system where the FDT is violated. The answer is no.
This provides  important  insight  not only   for the analysis of experimental data from NEQ systems but also
 for the systematic development
of methods from which the NEQ character of a system can be clearly identified.

To proceed,  the   Mori GLE for a general observable $A(t)$  reads
\cite{nakajima_quantum_1958,zwanzig_memory_1961,mori_transport_1965}
\begin{align} \label{eq_mori_GLE0}
 \ddot A(t)  = -  K    (A(t) - \langle  A \rangle)  
 - \int_{t_0}^{\infty} {\rm d}s\, \Gamma(t-s)  \dot A(s)  + F(t),
\end{align}
where the stiffness of the effective harmonic potential is denoted as $K$, 
$F(t)$ is the  complementary force
 and $\Gamma(t)$ is the single-sided  friction memory kernel, i.e. $\Gamma(t)=0$ for $t<0$. 
The time at which the projection onto the  initial probability distribution (taken to be
the canonical distribution)  is performed is denoted as $t_0$
(for details of the derivation see \cite{roland_neq_2023}).
Strictly speaking,  $F(t)$ is  a phase-space dependent  deterministic function
and  fulfills well-defined  initial conditions at $t=t_0$,
Eq. \eqref{eq_mori_GLE0} is thus  deterministic and fully  time reversible. 
As is usually done, 
we instead  treat $F(t)$ as a  random force characterized solely by  the two-point correlation 
 determined by
\begin{align} \label{eq_mori_GLE0b}
\langle F(s) F(t) \rangle =   \Gamma_R(t-s) =   \Gamma_R(s-t).
\end{align}
In EQ, the FDT dictates that 
$ \Gamma_R( t ) = \langle \dot A^2 \rangle  \Gamma(|t|)$,
in this paper we  treat the general case 
 $ \Gamma_R(t) \neq  \langle \dot A^2 \rangle \Gamma(|t|)$
that  is widely used as a model for NEQ systems.
In order to be able 
 to derive the  multi-point distribution explicitly,  the Mori 
  formulation of the GLE in Eq.\eqref{eq_mori_GLE0},
  which is linear in $A(t)$,  is crucial
  and absolutely sufficient for Gaussian observables as treated here.
We note in passing that
 the Mori GLE in Eq.\eqref{eq_mori_GLE0} is exact even for non-Gaussian observables, unless one approximates the random
 force distribution as Gaussian.

 
The question we address in this paper 
 is whether from a trajectory that follows from 
 the GLE defined by  Eqs. \eqref{eq_mori_GLE0} and  \eqref{eq_mori_GLE0b} 
 one can extract the GLE parameters, namely $K$, $\Gamma(t)$ and $ \Gamma_R(t)$.
 A more modest question is whether one is able to tell from the trajectory
 whether it comes from a NEQ GLE with 
  $ \Gamma_R(t) \neq  \langle \dot A^2 \rangle \Gamma(|t|)$. 
To answer both questions we derive the multi-point joint distribution for $A$  and show that 
it is determined solely in terms of the two-point correlation function. We also show 
that the two-point correlation function of the NEQ GLE with 
$ \Gamma_R(t ) \neq \langle \dot A^2 \rangle \Gamma(|t|)$
is identical to the two-point correlation function of an EQ GLE with
  $ \Gamma_R(t) / \langle \dot A^2 \rangle_{\rm eq} =   \Gamma(|t|) = 
\Gamma_{\rm eq} (|t|)$ with a uniquely determined single-sided effective EQ  kernel $\Gamma_{\rm eq} (t)$.
Since the multi-point distribution constitutes a complete
description of the dynamics of an observable,
by combining  these two results we conclude
that the  one-dimensional trajectory of a Gaussian NEQ observable   does not reveal its NEQ character.
This is a non-trivial and relevant finding, since many biological NEQ  observables,
such as the center-of-mass motion of cells \cite{Mitterwallner_2020}, 
can be described as  Gaussian processes  to high accuracy. 
  In this context it is important to realize
that an observable can be Gaussian even when  the entire many-body system is non-Gaussian, 
meaning that other observables and system coordinates may very well exhibit non-Gaussian fluctuations. 
So our results are not limited to Gaussian systems but only to Gaussian observables.

To proceed, 
by Fourier transforming Eq. \eqref{eq_mori_GLE0} according to 
$\tilde A(\nu)   = \int_{-\infty}^{\infty} dt e^{- \imath t \nu} A(t)$, 
we obtain the linear response relation as
\begin{align}  \label{eq_response}
\tilde A(\nu) - 2\pi \delta(\nu) \langle  A \rangle = \tilde \chi(\nu) \tilde F(\nu),
\end{align}
where the  Fourier-transformed  response function  (which is causal and thus single-sided in the 
time-domain) is determined  by 
\begin{align}  \label{eq_response2}
& 1 /  \tilde \chi(\nu) = K - \nu^2 + \imath \nu  \tilde \Gamma(\nu).
\end{align}
In deriving Eq. \eqref{eq_response} we have shifted the projection time into the far past, $t_0 \rightarrow - \infty$.
By doing so, the observable distribution and dynamics  become for systems with finite relaxation times (as we implicitly assume here)
 independent of the phase-space distribution imposed at the projection time $t_0$  when deriving
the GLE  \cite{roland_neq_2023}.

In order to derive the FDT  we  
 calculate the two-point correlation function 
\begin{align} \label{eq_corr}
C(t-t') =  \langle (A(t)-\langle A \rangle)(A(t')  -\langle A \rangle) \rangle
\end{align}
for general times $t,t' \geq t_0$  by  averaging over $F(t)$. 
By  combining   Eq.  \eqref{eq_corr} with  \eqref{eq_mori_GLE0b} and  \eqref{eq_response},
 the Fourier-transformed correlation function results as   \cite{Mitterwallner_2020} 
\begin{align}  \label{eq_corr3}
&\tilde C(\nu) = \tilde \chi(\nu) \tilde \chi(-\nu)  \tilde  \Gamma_R(\nu).
\end{align}
For an EQ  system with $\Gamma_R(t)/ \langle \dot A^2 \rangle_{\rm eq}  =\Gamma(|t|) \equiv  \Gamma_{\rm eq} (|t|)$
 the correlation function can be written  as 
\begin{align}  \label{eq_corr4}
&  \tilde C_{\rm eq} (\nu)   =  \tilde \chi_{\rm eq} (\nu) \tilde \chi_{\rm eq}(-\nu) \langle  \dot A^2 \rangle_{\rm eq}
 \left[   \tilde  \Gamma_{\rm eq}(\nu)  +   \tilde  \Gamma_{\rm eq}(-\nu)   \right]  \nonumber \\ 
&=  \langle  \dot A^2 \rangle_{\rm eq} \left[ 
 \frac{ \tilde \chi_{\rm eq}(-\nu)}{\imath \nu} - \frac{ \tilde \chi_{\rm eq}(\nu)}{\imath \nu} \right] 
 =- \frac{ 2  \langle  \dot A^2 \rangle_{\rm eq} \tilde \chi_{\rm eq}''(\nu)}{ \nu},
\end{align}
where the EQ response function is defined according to Eq.  \eqref{eq_response2} as 
$1/  \tilde \chi_{\rm eq} (\nu)   = K_{\rm eq}  - \nu^2 + \imath \nu  \tilde \Gamma_{\rm eq} (\nu)$.
In  contrast to the NEQ result in Eq. \eqref{eq_corr3}, $\tilde C_{\rm eq} (\nu)$  splits
into its two single-sided time-domain parts proportional to $\tilde \chi_{\rm eq}(\nu)$ 
and  $\tilde \chi_{\rm eq}(-\nu)$.
In the last step in Eq.  \eqref{eq_corr4}  we used that the  time-domain response function is real 
and thus 
$ \tilde \Gamma_{\rm eq}'(\nu)  = \tilde \Gamma_{\rm eq}' (-\nu)$ and 
$ \tilde \Gamma_{\rm eq}'' (\nu)  = -\tilde \Gamma_{\rm eq}'' (-\nu)$,
where we split $ \tilde \Gamma_{\rm eq}(\nu)$ into its real and imaginary parts
according to $ \tilde \Gamma_{\rm eq}(\nu)= \tilde \Gamma'_{\rm eq}(\nu) +
\imath \tilde \Gamma''_{\rm eq}(\nu)$.
Eq.  \eqref{eq_corr4}  is the  FDT  in the frequency domain and it obviously
only holds  in the EQ  case for  $\Gamma_R(t)/\langle  \dot A^2 \rangle_{\rm eq} =\Gamma(|t|)$. 
The mean square of the observable follows from  the correlation function as 
$ \langle A^2 \rangle_{\rm eq}  = C_{\rm eq}(0) = \int {\rm d} \nu  \tilde C_{\rm eq} (\nu)  = \langle  \dot A^2 \rangle_{\rm eq} / K_{\rm eq}$,
together with the relation
$ \langle \dot A^2 \rangle_{\rm eq}  =  - \ddot C_{\rm eq}(0) =
\int {\rm d} \nu \nu^2  \tilde C_{\rm eq} (\nu)$ 
this allows to determine  $K_{\rm eq}$ from   $\tilde C_{\rm eq} (\nu)$.

We next split the NEQ and EQ correlation functions into their single-sided components according
to $C(t)= C^+(t)+C^+(-t)$ and $C_{\rm eq}(t)= C_{\rm eq}^+(t)+C_{\rm eq}^+(-t)$,
where $C^+(t) \equiv  \theta(t) C(t)$ and $C_{\rm eq}^+(t) \equiv  \theta(t) C_{\rm eq}(t)$
and $\theta(t)$ denotes the Heavyside function. 
With this, the Fourier transform of 
$C_{\rm eq}^+(t)$  follows  from Eq.  \eqref{eq_corr4} 
  as
\begin{align}  \label{eq_corr5}
  \tilde C^+_{\rm eq} (\nu)   =  - 
  \frac{   \langle  \dot A^2 \rangle_{\rm eq}    }{\imath \nu} \left(  \tilde \chi_{\rm eq}(\nu) -  \tilde \chi_{\rm eq}(0) \right). 
\end{align}
Note that a similar decomposition  of Eq.  \eqref{eq_corr3}
is  not possible,   because  it  cannot
be straightforwardly  split into its  causal and anti-causal parts.

Now we ask whether for a given NEQ  correlation function 
$\tilde C(\nu)$ we can find a matching  EQ correlation function 
$\tilde C_{\rm eq}(\nu)$ that satisfies $\tilde C_{\rm eq}(\nu)= \tilde C(\nu)$
(in which case we would also have $ \langle  \dot A^2 \rangle_{\rm eq} =  \langle  \dot A^2 \rangle $
and $ \langle  A^2 \rangle_{\rm eq} =  \langle  A^2 \rangle$).
In fact, by combining the equivalent assumption for the single-sided correlation functions,
 $\tilde C^+_{\rm eq}(\nu) = \tilde C^+(\nu)$,
with Eq.  \eqref{eq_corr5} and the EQ version of Eq.  \eqref{eq_response2},
we find after  rearranging 
\begin{align}  \label{eq_corr6}
 \imath \nu  \tilde \Gamma_{\rm eq} (\nu) 
 =   \nu^2 -  K_{\rm eq}   +  \frac{ K_{\rm eq}  }{1 - \imath \nu   \tilde C^+ (\nu) / \langle  A^2 \rangle }, 
\end{align}
where $K_{\rm eq}=  \langle  \dot A^2 \rangle_{\rm eq} / \langle  A^2 \rangle_{\rm eq} 
=  \langle  \dot A^2 \rangle / \langle  A^2 \rangle$. In essence, Eq.  \eqref{eq_corr6} shows that
for every NEQ correlation function
that is produced  by the NEQ GLE in  Eqs.  \eqref{eq_mori_GLE0} and \eqref{eq_mori_GLE0b},
 there is  an EQ  GLE that displays the  identical correlation function, in other words, from the two-point
correlation function one can not decide whether a system is NEQ or EQ. At first sight, 
 the two-point correlation function seems like  only a partial description of a system, we therefore  ask next 
whether the general N-point distribution function allows to infer the NEQ properties of a system.

To proceed, we define  the $N$-point joint probability distribution of $A(t)$
 according to continuum probability calculus  as
\begin{align} \label{dist1}
 \rho \left ( \left \{A_N,t_N \right \}  \right ) 
=  \langle \prod_{k=1}^N  \delta (A_k-A(t_k) )  \rangle.
\end{align}
For the delta functions we use their  Fourier representation and obtain 
\begin{align} \label{dist2}
& \rho \left ( \left \{A_N,t_N \right \}  \right )  = 
\nonumber \\
&   \prod_{k=1}^N \int_{-\infty}^{\infty} \frac{{\rm d} q_k}{2\pi} 
 e^{\imath  \sum_{k=1}^N q_k A_k} 
\left  \langle \exp\left(  - \imath  \sum_{k=1}^N q_k A(t_k) \right) \right \rangle.
\end{align}
We next use the  linear response relation in Eq. \eqref{eq_response},
which is exact for a Gaussian observable,
and express the average over the random force $F(t)$ in Eq. \eqref{dist2}
 by a Gaussian path integral according to 
\begin{align}  \label{eq_resp2}
 & \langle  X(F)  \rangle =  \int_{-\infty}^{\infty} \frac{ {\cal D} F(\cdot)}{{\cal N}_F}
 X(F) \nonumber \\
 & \times  \exp\left( -  \frac{ 1}{ 2 } \int_{-\infty}^{\infty} {\rm d} s {\rm d} s' 
F(s) F(s') 
  \Gamma_R^{-1}(s-s') 
 \right),
 \end{align}
where ${{\cal N}_F}$ is a normalization constant
and $  \Gamma_R^{-1}(s)$ is  the  functional  inverse
of the random-force  kernels defined in Eq.  \eqref{eq_mori_GLE0b}
according to $\int {\rm d}s \Gamma_R^{-1}(t-s) \Gamma_R(s-t')= \delta(t'-t)$.
Since we consider a Gaussian variable $A(t)$, 
  non-Gaussian fluctuations of $F(t)$ 
  need   not  be considered here.
Performing the Gaussian path integrals, the $N$-point distribution follows in matrix
notation as
\begin{align} \label{dist3}
& \rho \left ( \left \{A_N,t_N \right \}  \right )  = 
\nonumber \\
&    \frac{  \exp \left(-  \sum_{j,k=1}^N (A_j -  \langle  A \rangle) I_{jk}^{-1}  (A_k -  \langle  A \rangle)/2  \right)}
 { \sqrt{ \det 2 \pi I}},
\end{align}
where the  matrix 
\begin{align} \label{dist4}
 I_{jk} = C(t_j - t_k)
\end{align}
 is determined in terms of the two-point correlation function defined in Eq. \eqref{eq_corr3},
 details of the derivation are given in the  Supplemental Material  \cite{SM}.
From Eq.  \eqref{dist3} we see that if the  random force $F(t)$ is  described by a 
Gaussian process, as assumed in Eq. \eqref{eq_resp2},
then also the observable $A(t)  - \langle  A \rangle $
is a Gaussian process determined by the correlation function $C(t)$ defined in  Eq. \eqref{eq_corr3}.
Since the correlation function $C(t)$ is identical to an EQ correlation function 
$C_{\rm eq}(t)$ with a memory kernel defined in Eq.  \eqref{eq_corr6},
and since the $N$-point joint  distribution for general $N$  completely defines the trajectory, 
we conclude that there is no way to tell whether a process is NEQ or EQ from 
a one-dimensional trajectory of a general observable. Note that distributions
involving velocities and higher-order derivates can be constructed from the positional 
$N$-point distribution in Eq.  \eqref{dist3} by suitable contraction,
so our conclusion also holds if one considers more general multi-point joint distributions involving
derivatives of the observable. 

While the N-point distribution of a Gaussian NEQ trajectory does not reveal its NEQ character,
it can very well be used to characterize its non-Markovianity. To make this notion concrete,
we consider the three-point distribution function 
$ \rho  ( A_3,t_3 ;A_2, t_2; A_1,t_1)$, which according to Eq.  \eqref{dist3}
 is determined  by the inverse covariance matrix
 
  \begin{equation} \label{matrix}
I^{-1}  = \frac{\begin{pmatrix}
 J^2 -I^2_{23}			& I_{13} I_{23} - J  I_{12}  	&  I_{12} I_{23} - J I_{13}  \\
 I_{13} I_{23} - J  I_{12}    	& J^2 -I^2_{23}			 &  I_{12} I_{13} - J I_{23} \\
 I_{12} I_{23} - J I_{13} 	 &  I_{12} I_{13} - J  I_{23}	 &  J^2 -I^2_{12}
\end{pmatrix}} {\det I},
\end{equation}
where  we introduced the short-hand notation  $J\equiv C(0)=\langle A^2 \rangle $ 
for the mean square of $A$. 
There are different ways of defining the Markovian property, we here use the 
factorization 
of the three-point conditional distribution  into the product of two two-point conditional
distributions according to  \cite{zwanzig_nonequilibrium_2001}
\begin{align}  \label{dist5}
\rho  ( A_3, t_3 ; A_2, t_2 |  A_1,t_1) = \rho  ( A_3, t_3  | A_2, t_2 )\rho  ( A_2, t_2 |  A_1,t_1),
\end{align}
where the conditional three- and two-point distributions are defined  as
$\rho  ( A_3, t_3 ; A_2, t_2 |  A_1,t_1)  \equiv  \rho  ( A_3,t_3 ;A_2, t_2; A_1,t_1) / \rho(A_1,t_1)$
and  $ \rho  ( A_3, t_3  | A_2, t_2 ) \equiv  \rho  ( A_3, t_3  ; A_2, t_2 ) / \rho(A_2,t_2)$
and  the single-point distribution  follows from  Eq. \eqref{dist3} as
$ \rho(A_1,t_1) = \exp(- (A_1 - \langle A \rangle )^2 /(2 J)) / \sqrt{ 2 \pi J}$ 
and obviously is independent of time. 
It is easy to see that the Markovian condition Eq. \eqref{dist5}
can only be satisfied if the off-diagonal entries $I^{-1}_{13} = I^{-1}_{31}$ 
of the inverse covariance matrix in Eq.  \eqref{matrix} vanish, i.e.,  if 
$ I_{12} I_{23} - J I_{13} =0$ holds (this condition is in fact necessary and sufficient). 
Using the definition Eq. \eqref{dist4} the Markovian condition can thus be written as
\begin{align}  \label{dist6}
C(t_3-t_1) = C(t_3-t_2) C(t_2 - t_1)/C(0),
\end{align}
which is only satisfied if the  correlation function  is a single exponential,
$C(t)= C(0)e^{-t/\tau}$, as is well known. 
We conclude that the only trajectory property  one can deduce from the multi-point distribution
is the non-Markovianity of the observable 
by the violation of Eq.  \eqref{dist5} (which strictly speaking 
is equivalent to  deviations of the two-point correlation function $C(t)$
from a single exponential decay), but not the NEQ character of the observable.
To deduce the NEQ properties one either needs to observe more than one observable,
as has been reviewed  recently \cite{BroederszReview}, 
or to simultaneously measure correlations and the response to an externally applied force
\cite{Schmidt07,Netz2018}.
In the following we present an alternative way that involves switching  the system from EQ into 
NEQ at a well-defined time,
which allows to extract all  parameters  of the  NEQ GLE.

The standard way of extracting the memory kernel from time-series data is by turning the stochastic  
GLE for the observable  $A(t)$  into a non-stochastic integro-differential 
equation for the two-point correlation function,  which can be  solved by Fourier transform or 
 recursively after discretization \cite{Straub_1987}.
Here we show that the same recipe also works for the
NEQ  GLE. To proceed, we multiply the GLE in Eq. \eqref{eq_mori_GLE0} by 
$\dot A( t_0)$, the velocity at the projection time $t_0$,  and average over the  random  force, 
by which we obtain
\begin{align}  \label{eq_Volt1}
&- \dddot C_0(t-t_0) = K \dot C_0(t-t_0) +\int_0^{t-t_0}  {\rm d}s 
\Gamma(s) \ddot C_0(t-t_0-s) 
\end{align}
for the two-point correlation function 
\begin{align} \label{eq_Volt2}
C_0(t-t_0) = \langle (A(t_0)-\langle A \rangle)(A(t)  -\langle A \rangle) \rangle.
\end{align}
Note that the two-point correlation function $C_0(t)$ defined here differs from the one defined in 
Eq. \eqref{eq_corr} in that  one of the times coincides with  the projection time $t_0$.
This makes a fundamental difference, as will become clear shortly. 
When deriving Eq. \eqref{eq_Volt1},  we used  that the  average $\langle \dot A(t_0) F(t)\rangle$
vanishes \cite{roland_neq_2023}.
Performing a single-sided Fourier transform of  Eq. \eqref{eq_Volt1} 
and defining   $\tilde C_0^+(\nu)   = \int_{0}^{\infty} dt e^{- \imath t \nu}C_0(t)$, we obtain
the solution
\begin{align}  \label{eq_Volt3}
\tilde C_0^+(\nu)  = \frac{\ddot C_0(0) \tilde \chi(\nu)}{\imath \nu} + \frac{C_0(0)}{\imath \nu} 
\end{align}
in terms of the response function $ \tilde \chi(\nu)$ defined in Eq. \eqref{eq_response2}.
In other words, from the time-domain correlation function $C_0(t)$, which can be 
straightforwardly  obtained in experiments or simulations 
by turning on the NEQ-producing conditions  at  time $t_0$ 
(assuming that the distribution up to $t_0$   equals
the canonical projection distribution), 
 the single-sided Fourier transform $\tilde C_0^+(\nu)$
and the values $\ddot C_0(0)$ and $C_0(0)$ follow, from  which $\tilde \chi(\nu)$ is determined via
Eq. \eqref{eq_Volt3} by direct  inversion. 
 Alternatively, instead of Fourier transformation, 
Eq. \eqref{eq_Volt1} can be recursively   solved by discretization  \cite{carof_two_2014,Ayaz2022}. 
We conclude that from the correlation functions $C(t)$ and $C_0(t)$, defined in Eqs. 
 \eqref{eq_corr} and \eqref{eq_Volt2}, all parameters of the GLE can be determined via
 Eqs.   \eqref{eq_corr3} and \eqref{eq_Volt3} 
 and in particular
 the NEQ character of the observable can be deciphered. 

In conclusion, by simply observing the trajectory of a one-dimensional Gaussian observable over time,
it cannot be decided whether the observable characterizes a NEQ or an EQ system, regardless
of how the trajectory is processed or  manipulated. In contrast,  the non-Markovianity
of an observable easily follows from the trajectory by using  standard criteria.
 For the GLE studied here, NEQ can only be obtained for non-vanishing memory, i.e. if the system
is non-Markovian, because for a Markovian system the FDT can trivially be restored by redefinition
of the temperature. But, obviously, the opposite is not true, not every non-Markovian system is necessarily NEQ.
Thus, non-Markovianity and NEQ are observable properties that are distinct but impossible to disentangle solely based
on trajectory information.
These findings also hold for non-Gaussian systems,  as we did not assume
that the entire system is Gaussian, but rather     only the considered observable. 
As a matter of fact,
this is true for cancer cells, the motion of which  has been recently demonstrated
to be perfectly Gaussian when analyzed on a single-cell level \cite{Mitterwallner_2020}.
Thus, to deduce the NEQ properties of a system  one either needs to simultaneously observe more than one observable
 \cite{BroederszReview} or one needs to apply an external force and to explicitly detect the violation of the FDT
\cite{Schmidt07,Netz2018}.
We here  present an alternative way that involves switching on the NEQ character  of the system at a well-defined time
and to measure two-point correlation functions both at the switching time  and at later times.

Very recently it has been shown that for a model defined by  a time-dependent Hamiltonian 
that includes an external NEQ force  which couples linearly to an observable, the resulting 
GLE nevertheless obeys the FDT if the observable described by the GLE is  Gaussian \cite{roland_neq_2023}.
Our results reported here are in line with those findings  but constitute  a considerable generalization,
as no explicit form of the underlying NEQ Hamiltonian needs to be assumed  for the present derivations.



\begin{acknowledgments}
We acknowledge support by Deutsche Forschungsgemeinschaft Grant CRC 1114 "Scaling Cascades in Complex System", 
Project 235221301, Project B03, and 
by the ERC Advanced Grant 835117 NoMaMemo
and by the Infosys Foundation. 
\end{acknowledgments}

\bibliography{NonEq}

\end{document}